\documentclass[12pt]{article}
\usepackage{a4wide,epsfig,amsmath,amssymb,cite}
\usepackage{scalefnt}
\newcommand{\yh}{y_H}

\newcommand{\yz}{y_Z}

\newcommand{\yhonea}{y_{H,1a}}
\newcommand{\yhoneb}{y_{H,1b}}

\newcommand{\yzonea}{y_{Z,1a}}
\newcommand{\yzoneb}{y_{Z,1b}}

\begin{document}


\title{\vskip-3cm{\baselineskip14pt
\begin{flushleft}
\normalsize SFB/CPP-06-09 \\
\normalsize TTP06-09 \\
\end{flushleft}}
\vskip1.5cm
Complete Higgs mass dependence of top quark pair threshold production to
order $\alpha\alpha_s$}
\author{\small D. Eiras and M. Steinhauser\\
  {\small\it Institut f{\"u}r Theoretische Teilchenphysik,
    Universit{\"a}t Karlsruhe}\\
  {\small\it 76128 Karlsruhe, Germany}\\
}

\date{}

\maketitle

\thispagestyle{empty}

\begin{abstract}
In this paper we consider the production of top quark pairs close to
threshold and compute the dependence on the Higgs boson mass to
order $\alpha\alpha_s$.
This requires the evaluation of the matching coefficient
of the vector current to two-loop level and the inclusion of  
Higgs mass dependent operators in the non-relativistic effective theory.
For Higgs boson masses below 200~GeV moderate contributions to the top
quark mass and the peak cross section are observed.
We also provide additional information to the on-shell
wave function renormalization which is relevant for the
matching coefficient.
\medskip

\noindent
PACS numbers: 14.65.Ha, 14.80.Bn
\end{abstract}

\newpage


\section{Introduction}

One of the most important tasks of a future international linear
collider (ILC)~\cite{ilc} is the precise measurement of the cross section
for the production of top quark pairs close to threshold.
The comparison with theoretical calculations will lead to a
determination of the top quark mass with an unrivaled
precision leading to an uncertainty below 100~MeV~\cite{Martinez:2002st}. 
Furthermore, also the width of the top quark, the strong
coupling and the Yukawa coupling between the top quark and Higgs boson can
be extracted from such a measurement.
The findings of this paper are important in this context.
However, the final success of such an enterprise depends crucially on 
whether the theoretical precision can match the expected experimental 
one. The complete next-to-next-to-leading analysis of
Ref.~\cite{Hoang:2000yr} has shown that it is important to obtain the 
third-order result within QCD. This constitutes a long-term 
calculation which has already been started (see, e.g.,
Refs.~\cite{Penin:2005eu,Beneke:2005hg} for the most recent results). 

The naive scaling rule $\alpha \sim \alpha_s^2$,
where $\alpha$ is the fine structure and $\alpha_s$ is the strong
coupling constant, shows that the next-to-next-to-next-to-leading
terms are parametrically of the same order as the mixed 
corrections proportional to $\alpha\alpha_s$. 
In this paper we take the first step to a complete order
$\alpha\alpha_s$ result and
compute the full dependence on the Higgs boson
mass. On one hand this requires the
evaluation of two-loop vertex corrections within the Standard Model (SM)
involving a Higgs boson and a gluon. Furthermore, a new operator
depending on the Higgs mass has to be considered on the effective theory side.
As we will explicitly see, the two individual pieces are divergent,
however, the physical quantities formed by the proper combination is 
finite.

Effects from the Higgs boson mass have already been studied in
Ref.~\cite{Strassler:1990nw} where a Yukawa-type potential has been
used and the Schr\"odinger equation has been solved numerically in
order to obtain the imaginary part of the Green function and finally
the total cross section. In this paper we will consider the Higgs
boson mass to be much larger than the soft scales involved in the
process and include it in a systematic way in the non-relativistic
effective theory.
The importance of a systematic and careful treatment of electroweak
corrections to the top quark pair production has been stressed in
Ref.~\cite{Hoang:2004tg,Hoang:2006pd}, where certain 
next-to-next-to-leading logarithmic electroweak effects associated
to the instability of the top quark have been considered.

The paper is organized as follows: In the next Section we consider 
the new Higgs mass dependent operator included into non-relativistic
QCD (NRQCD) and evaluate the
corrections to the energy level and the wave function at the origin.
In Section~\ref{sec::cv} the Higgs mass dependent corrections to the
matching coefficient are discussed and Section~\ref{sec::appl}
contains a brief account of the phenomenological applications of our
results. We conclude in Section~\ref{sec::concl}.
In Appendix~\ref{app::Z-boson} we discuss the results for the matching
coefficients induced by vertex corrections involving a $Z$ boson and 
Appendix~\ref{app::z2} contains
additional information on the wave function renormalization constant
which is relevant in connection to the matching coefficient.


\section{Effective theory}

The standard theoretical framework for the
threshold production of top quark pairs is given by 
NRQCD~\cite{Caswell:1985ui,Bodwin:1994jh} 
which ensures
the automatic resummation of terms like $(\alpha_s/v)^n$
($n=1,2,3,\ldots$) where $v$ is the velocity of the produced
quarks. NRQCD is constructed from QCD by integrating out the hard
scales associated to the mass of the heavy quark, $m$.
Thus it contains only degrees of freedom of the order $mv$ and $mv^2$.

In this paper we consider next to the top quark, which takes over the role
of the heavy quark,
also the Higgs boson with mass $M_h$. 
Since $M_h\gg m_t v$ both mass scales are integrated out
simultaneously.
In Ref.~\cite{Kniehl:2002br} all operators
which appear within QCD up to third order in perturbation theory 
have been classified.
The only operator which has to be added to the Hamiltonian in Eq.~(6)
of Ref.~\cite{Kniehl:2002br} is given by
\begin{eqnarray}
  \delta {\cal H}_H &=& - \frac{\alpha \pi m_t^2}{s_W^2 M_W^2 M_H^2} 
  \,,
  \label{eq::Hop}
\end{eqnarray}
where $s_W$ is the sine of the Weinberg angle and $M_W$ is the $W$
boson mass. The corresponding expression in coordinate space is
proportional to the delta function.
If we employ the counting rule $\alpha \sim \alpha_s^2$ it is easy to
see that $\delta {\cal H}_H$ gives contributions which are
parametrically of the same order as the ones from the third-order QCD 
Hamiltonian.

Let us in a first step evaluate the corrections to the energy levels
induced by the operator $\delta {\cal H}_H$. Using first order
perturbation theory we obtain
\begin{eqnarray}
  \delta E_n^H &=& \langle\psi_n^C|\delta {\cal H}_H|\psi_n^C\rangle
  \,\,=\,\,
  E_n^C \frac{\alpha\alpha_s C_F m_t^4}{2 s_W^2 n M_W^2 M_h^2}
  \,,
  \label{eq::E_n^H}
\end{eqnarray}
with $E_n^C = - C_F^2\alpha_s^2 m_t/(4 n^2)$ and $\psi_n^C$ is the
Coulomb wave function. $n$ is the principal quantum number.

In order to compute the correction to the wave function at the
origin the operator of Eq.~(\ref{eq::Hop}) has to be inserted into the
standard formulae of non-relativistic perturbation theory
\begin{eqnarray}
  \delta\psi_n(0) = -\int {\rm d}^3{\vec{r}}\, 
  G_C(0,{\vec{r}\,},E) \, \delta {\cal H}_H \, \psi^C_n({\vec{r}})
  \,,
  \label{eq::dpsi_c}
\end{eqnarray}
where $G_C$ is the Coulomb Green function. Following
Ref.~\cite{Kniehl:2002yv} we split $G_C$ into a contribution with
zero, one and infinitely many gluon exchanges. 
Since only the one-gluon-exchange part is divergent it is convenient
to perform the corresponding calculation in momentum space whereas the
other contributions are evaluated in coordinate space. As a final
result we obtain
\begin{eqnarray}
  \delta \psi_n^H(0) &=& \psi^C_n(0)
  \frac{\alpha\alpha_s m_t^4}{s_W^2 M_W^2 M_h^2} C_F \left[
    - \frac{1}{4} \ln\left(\frac{\alpha_s C_F m_t}{\mu}\right)
    + \frac{3}{8}
    \right]
  \,,
  \label{eq::psin}
\end{eqnarray}
with 
$\left|\psi^C_n(0)\right|^2=C_F^3\alpha_s^3m_t^3/(8\pi n^3)$.
The divergence in Eq.~(\ref{eq::psin}) has been subtracted minimally
which will also be done for the coefficient function considered in the
next section.

We want to mention that the formulae of this Section are adapted for
the top quark. However, they also apply to other
quark masses, in particular to the bottom system,
by simply exchanging the top quark mass.


\section{\label{sec::cv}Higgs mass dependence of the matching coefficient}

The matching coefficient of the vector current 
$j^\mu = \bar{t} \gamma^\mu t$ is defined by
\begin{eqnarray}
  j^k_v = c_v \phi^\dagger \sigma^i \chi + {\cal
  O}\left(\frac{1}{m_t^2}\right)
  \label{eq::def_of_cv}
  \,,
\end{eqnarray}
where $\phi$ and $\chi$ are two-component Pauli spinors for quark and
anti-quark, respectively, and
$k=1,2,3$ denote the spacial components. 
Note that there is no contribution to our order from the time-component.

For the practical computation of $c_v$ 
it is convenient to consider the $t\bar{t}\gamma$ vertex in the limit
where for the photon energy, $s$, we have $s\approx 4 m_t^2$.
In this case we can establish the equation
\begin{eqnarray}
  Z_2 \Gamma_v &=& c_v \tilde{Z}_2 \tilde{Z}_v^{-1} \tilde{\Gamma}_v
  + \ldots
  \,,
  \label{eq::def_cv}
\end{eqnarray}
where we have on the left- and right-hand side quantities of the full and
effective theory, respectively. The latter are marked
by a tilde and the ellipses denote terms suppressed by inverse powers
of the top quark mass. 
$\Gamma_v$ denotes the $t\bar{t}\gamma$ 
vertex corrections where it is understood that
the couplings and masses are renormalized.
The two-loop mixed correction to the on-shell wave function
renormalization, $Z_2$, has been computed recently in 
Ref.~\cite{Eiras:2005yt}. 
In particular, for the Higgs boson contributions both 
the exact result and the expansions in three kinematic regions have
been provided.

At this point it is convenient to apply the so-called threshold
expansion~\cite{Beneke:1997zp,Smirnov:pj} to Eq.~(\ref{eq::def_cv})
which has the consequence that $\Gamma_v$ has to be evaluated for 
$s=4m_t^2$ since all except the hard region cancel in Eq.~(\ref{eq::def_cv}).
Furthermore, on the right-hand side only tree contributions have to be
considered. In particular we have $\tilde{Z}_2=1$.

The one-loop corrections to $c_v$ are finite~\cite{Guth:1991ab}. However,
starting from two-loop order, the matching coefficient
exhibits infra-red divergences which are compensated by ultra-violet
divergences of the effective theory rendering physical quantities
finite. In Eq.~(\ref{eq::def_cv}) the renormalization constant
$\tilde{Z}_v$ which generates the anomalous dimension of $\tilde{j}_v$
takes over this part. 
Note that the vector current in the full theory does not get
renormalized. 

It is convenient to introduce the perturbative decomposition of the
matching coefficient
\begin{eqnarray}
  c_v &=& 1 + \frac{\alpha_s}{\pi} C_F c_v^{\rm QCD,1}
  + \left(\frac{\alpha_s}{\pi}\right)^2 C_F c_v^{\rm QCD,2}
  + \frac{\alpha}{\pi s_W^2} c_v^{\rm ew}
  + \frac{\alpha\alpha_s}{\pi^2 s_W^2} C_F c_v^{\rm mix}
  \,,
  \label{eq::cvdef}
\end{eqnarray}
where analog formulae also hold for $Z_2$ and $\Gamma_v$.
Since we consider in this paper only the contribution 
from the Higgs boson the corresponding quantities obtain an additional
superscript $H$.

The two-loop QCD corrections, $c_v^{\rm QCD,2}$, have been computed 
in Refs.~\cite{Czarnecki:1997vz,Beneke:1997jm} (see also
Ref.~\cite{Kniehl:2006qw}). 
The complete SM corrections to one-loop order 
can be found in Ref.~\cite{Guth:1991ab}
where next to the vertex corrections also the box diagrams
contributing to $e^+ e^- \to t\bar{t}$ have been considered.
The latter, however, get no contributions from the Higgs boson.

From the perturbative expansion of Eq.~(\ref{eq::def_cv}) it is easy
to see that the one-loop results $c_v^{\rm QCD,1}$ and $c_v^{\rm ew}$
are simply given by the sum of the one-loop expressions for $Z_2$ and
$\Gamma_v$. Whereas the individual pieces are still divergent the
proper sum is finite. For convenience we repeat the QCD corrections and
the Higgs mass dependent term of $c_v^{\rm ew}$ which are given by
\begin{eqnarray}
  c_v^{\rm QCD,1} &=& -2\,,
  \nonumber\\
  c_v^{H, \rm ew} &=& \frac{m_t^2}{M_W^2}
  \Bigg[ \frac{3\yh^2-1}{12\yh^2}-\frac{2-9\yh^2+12\yh^4}{48\yh^4}
  \ln\yh^2 - \frac{(-2+5\yh^2-6\yh^4)}{24\yh^2}\Psi(\yh) \Bigg]\, ,
  \label{eq::cv1exact}
\end{eqnarray}
where
\begin{eqnarray}
 \Psi(x)& = &\left\{
 \begin{array}{ll} 
   \frac{\sqrt{4x^2-1}}{x^2}
   \arctan{\sqrt{4x^2-1}} & \mbox{for} \,\, x\ge\frac{1}{2}  
   \\ [1em]
   \frac{\sqrt{1-4x^2}}{2 x^2}
   \ln{\frac{1-\sqrt{1-4x^2}}{1+\sqrt{1-4x^2}}} & \mbox{for} \,\,
   x<\frac{1}{2} 
   \,,
 \end{array} 
 \right.
 \label{eq::psi}
\end{eqnarray}
and $\yh=m_t/M_h$.

At two-loop order it is quite difficult to obtain a closed
analytic result valid for all Higgs and top quark masses.
However, it is possible to get compact formulae valid in
various kinematical regions which --- when combined --- cover the
whole Higgs mass range. This strategy has been successfully applied in
Ref.~\cite{Eiras:2005yt} to the on-shell wave function renormalization
(see also Appendix~\ref{app::z2}). 
Thus, let us consider $c_v^{H, \rm ew}$ in the limits 
$m_t\ll M_H$, $m_t\approx M_H$ and $m_t\gg M_H$
where it is given by
\begin{eqnarray}
  c_{v,0}^{H, \rm ew} &=& \frac{m_t^2}{M_W^2}\Bigg[
    \yh^2 \left(\frac{31}{144} -
    \frac{5}{24}\ln\yh^2
    \right)
    + \yh^4\left(-\frac{3}{16} - \frac{\ln\yh^2}{4}\right) 
    + \yh^6\left(-\frac{307}{480} - \frac{5\ln\yh^2}{8}\right)  
    \nonumber\\ &&\mbox{} 
    + \yh^8\left(-\frac{737}{360} - \frac{11\ln\yh^2}{6}\right) 
    + \ldots
    \Bigg]
  \,,\nonumber\\
  c_{v,1a}^{H, \rm ew} &=& \frac{m_t^2}{M_W^2}\Bigg[
    \frac{1}{6} + \frac{\pi}{8\sqrt{3}} + \yhonea \left (\frac{1}{24}
    + \frac{\pi}{8\sqrt{3}}\right) + \yhonea^2\left(-\frac{1}{12} +
    \frac{\pi}{6\sqrt{3}}\right) 
    \nonumber\\&&\mbox{}  
    + \yhonea^3 \left (-\frac{1}{36} + \frac{13\pi}{108\sqrt{3}}\right)+
    \yhonea^4\left(\frac{1}{288} + 
    \frac{5\pi}{54\sqrt{3}}\right)+ \ldots \Bigg]
  \,,\nonumber\\
  c_{v,1b}^{H, \rm ew} &=& \frac{m_t^2}{M_W^2}\Bigg[
    \frac{1}{6} + \frac{\pi}{8\sqrt{3}} - 
    \yhoneb \left( \frac{1}{24} + \frac{\pi}{8\sqrt{3}}\right) 
    + \yhoneb^2 \left( -\frac{1}{8} + \frac{\pi}{24\sqrt{3}} \right )  
    \nonumber\\ &&\mbox{} 
    - \yhoneb^3 \left ( \frac{13}{72} - \frac{19\pi}{216\sqrt{3}}
    \right ) + \yhoneb^4 \left(-\frac{59}{288} +
    \frac{23\pi}{216\sqrt{3}} \right ) +\ldots  
    \Bigg]
  \,,\nonumber\\
  c_{v,\infty}^{H, \rm ew} &=& \frac{m_t^2}{M_W^2}\Bigg[
      \frac{\pi}{4}\yh
    - \frac{\ln\yh}{2}
    - \frac{23\pi}{96}\frac{1}{\yh} + \left(\frac{7}{48} +
    \frac{3}{8}\ln\yh\right)\frac{1}{\yh^2}  
    + \frac{55\pi}{512}\frac{1}{\yh^3} 
    \nonumber \\&& 
    + \left(-\frac{71}{720} -
    \frac{\ln\yh}{12}\right)\frac{1}{\yh^4} + \ldots 
  \Bigg]
    \,,
    \label{eq::cv1exp}
\end{eqnarray}
with $\yhonea = (1-1/\yh^2)$ and $\yhoneb = (1-\yh^2)$.
$c_{v,1a}^{H, \rm ew}$ and $c_{v,1b}^{H, \rm ew}$ are two different
representations of the same information which turn out to be useful
for different values of $\yh$.

\begin{figure}[t]
  \begin{center}
    \begin{tabular}{c}
      \epsfig{figure=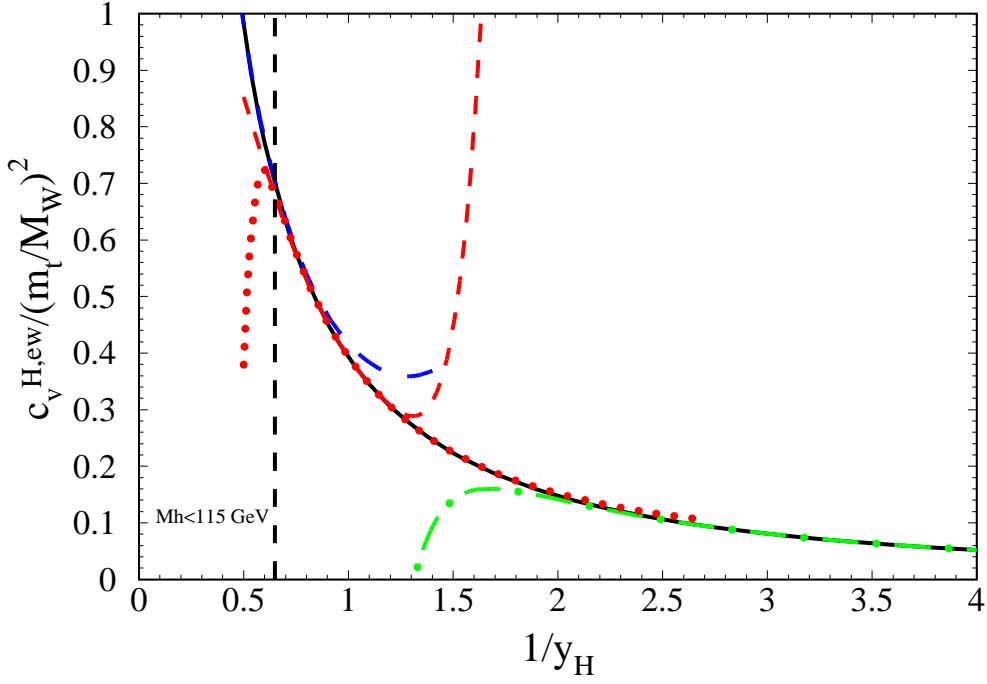,width=35em}
    \end{tabular}
    \parbox{14.cm}{
      \caption[]{\label{fig::cv1l}\sloppy
        $c_{v}^{H, \rm ew}/(m_t^2/M_W^2)$ as a function of $1/\yh$.
        The solid (black) line represents the exact result.
        The approximations in the three regions are shown as 
        dotted, dashed and dash-dotted lines where for the expansion
        around $\yh\approx 1$ two different approximations have been
        chosen. 
        }}
  \end{center}
\end{figure}

In Fig.~\ref{fig::cv1l} the exact result for $c_v^{\rm ew}/(m_t^2/M_W^2)$
(full black line, cf. Eq.~(\ref{eq::cv1exact})) 
is shown together with the expansions from Eq.~(\ref{eq::cv1exp}),
$c_{v,0}^{H, \rm ew}$ (dash-dotted), $c_{v,1a}^{H, \rm ew}$ (short dashed),
$c_{v,1b}^{H, \rm ew}$ (dotted) and $c_{v,\infty}^{H, \rm ew}$
(long dashed).
One can see that the approximations nicely cover the whole region of
$\yh$. 
Note that the expansion around $m_t=M_H$ shows better convergence
properties in case $\yhoneb$ is used as parameter.

Let us draw the attention of the reader to the expansion terms
for large top quark masses which starts with an enhancement 
factor $m_t^3/(M_W^2 M_h)$. 
Two out of the three powers in $m_t$ come from the Yukawa coupling
between the Higgs boson and the top quark and the factor $m_t/M_h$ is
the indication of the Coulomb singularity which would be present 
for a massless Higgs boson.
The next-to-leading term is quadratic in $m_t$ accompanied by a
logarithm in $m_t/M_h$. Both terms indicate that even for the leading terms
it is not possible to nullify the Higgs boson mass and non-trivial integration
regions have to be considered which makes the evaluation of the
corresponding expansion terms quite tedious. This is particularly true
for the two-loop order where due to the Coulomb divergence
one will have a further factor $m_t/M_h$ as compared to the
one-loop term. Thus the expansion starts with a quartic top quark mass
dependence. Furthermore, there are momentum regions which have
$\sqrt{m_t/M_H}$ as expansion parameter.

On the other hand, as can be seen in
Fig.~\ref{fig::cv1l}, the result from this region is
phenomenologically less important since for Higgs boson masses above
$115$~GeV there is perfect agreement between the exact result and the
approximation one obtains for $m_t\approx M_h$ and $m_t\ll M_h$.
Higgs boson masses below approximately $115$~GeV are excluded by the
direct search at the CERN Large Electron Positron Collider (LEP).
For this reason we do not consider the limit $m_t\gg M_h$ at two-loop order.

\begin{figure}[t]
  \begin{center}
    \begin{tabular}{c}
    \epsfig{figure=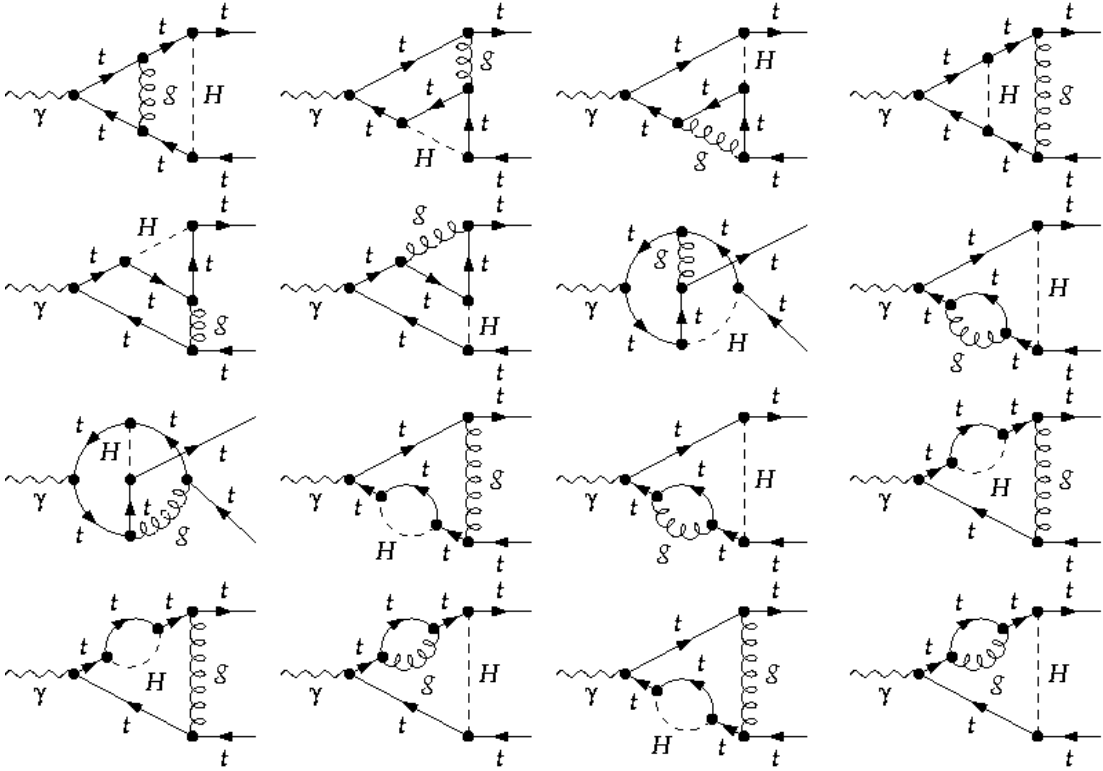,width=36em}
    \end{tabular}
    \parbox{14.cm}{
      \caption[]{\label{fig::diags}\sloppy Two-loop Feynman diagrams
        contributing to $c_v^{H,\rm mix}$.
        }}
  \end{center}
\end{figure}

Let us now turn to two loops. The Feynman diagrams are shown in
Fig.~\ref{fig::diags}.
They are generated with {\tt QGRAF}~\cite{Nogueira:1991ex}.
The application of {\tt q2e} and 
{\tt exp}~\cite{Harlander:1997zb,Seidensticker:1999bb}
identifies the topology of the individual diagrams
and adopts the notation in order to match one of the following
functions
\begin{eqnarray}
  \lefteqn{J^\pm(n_1,n_2,n_3,n_4,n_5) =}\nonumber\\&&
    \frac{e^{2\epsilon\gamma_E}}{(i\pi^{d/2})^2} 
  \int\frac{ {\rm d}^d k {\rm d}^d l }{(k^2+2kq)^{n_1}
    (l^2\pm 2lq)^{n_2}(k^2)^{n_3}((k-l)^2)^{n_4}(l^2-M^2)^{n_5}}
  \,,
  \nonumber\\
  \lefteqn{H^\pm_N(n_1,n_2,n_3,n_4,n_5) =}\nonumber\\&&
  \frac{e^{2\epsilon\gamma_E}}{(i\pi^{d/2})^2} 
  \int\frac{ {\rm d}^d k {\rm d}^d l }{(k^2+2kq)^{n_1}
    (l^2\pm 2lq)^{n_2}(k^2)^{n_3}((k-l)^2-M^2)^{n_4}(l^2)^{n_5}}
  \,,
  \nonumber\\
  \lefteqn{Y^\pm_N(n_1,n_2,n_3,n_4,n_5) =}\nonumber\\&&
  \frac{e^{2\epsilon\gamma_E}}{(i\pi^{d/2})^2} 
  \int\frac{ {\rm d}^d k {\rm d}^d l }{(k^2+2kq)^{n_1}
    (l^2 \pm 2lq)^{n_2}((k-l)^2 \mp 2q(k-l))^{n_3}(k^2)^{n_4}(l^2-M^2)^{n_5}}
  \,,
  \nonumber\\
  \lefteqn{Z^+_N(n_1,n_2,n_3,n_4,n_5) =}\nonumber\\&&
    \frac{e^{2\epsilon\gamma_E}}{(i\pi^{d/2})^2} 
  \int\frac{ {\rm d}^d k {\rm d}^d l }{(k^2+2kq)^{n_1}
    (l^2 + 2lq)^{n_2}((k-l)^2 - 2q(k-l))^{n_3}(k^2-M^2)^{n_4}(l^2)^{n_5}}
  \,.
  \label{eq::FDclasses}
\end{eqnarray}
where $d=4-2\epsilon$ is the space-time dimension and the $n_i$ are
integer indices. The choice of the five-line integrals of
Eq.~(\ref{eq::FDclasses}) is possible due to the special kinematics we
have at hand. Note that $J^+$, $H^+$ and $Y^-$
correspond to self energies whereas $J^-$, $H^-$, $Y^+$ and $Z^+$ to vertex
diagrams. 

In a next step we use the program {\tt AIR}~\cite{Anastasiou:2004vj}
in order to reduce the integrals to about 30 
master integrals. They range from very simple two-point expressions up
to complicated two-scale integrals with five lines. 
At this point an asymptotic expansion in the various
kinematical regions is performed.
From the one-loop result (cf. Fig.~\ref{fig::cv1l}) and from the
considerations in the context of the on-shell wave function
renormalization (see Ref.~\cite{Eiras:2005yt} and Appendix~\ref{app::z2}) 
one can see that for the phenomenological interesting Higgs boson
masses it is sufficient to have the
expansion around $m_t\approx M_H$ and the one for $m_t\ll M_H$
at hand.
Thus the master integrals are expanded in these two
limits.
In the case $m_t\approx M_H$ this reduces to a simple Taylor
expansion whereas for $m_t\ll M_H$ the well-established hard-mass
procedure~\cite{Smirnov:pj} is applied. The latter is actually
automated in the 
program~{\tt exp}~\cite{Harlander:1997zb,Seidensticker:1999bb}.
Hence as an independent check we applied with the help of {\tt exp} the
hard-mass procedure to each diagram which immediately leads to simpler
expressions and avoids the reduction to the master integrals.
In this way the result for $m_t\ll M_h$ could be checked.
We want to mention that the calculation has been performed for general
QCD gauge parameter, $\xi$. The cancellation of $\xi$ in the final
result serves as a welcome check for our calculation.

As already mentioned in the Introduction, even after the proper
combination of $Z_2$ and $\Gamma_v$ is formed, as prescribed by
Eq.~(\ref{eq::def_cv}), there remains an infra-red divergence which is
plugged into $\tilde{Z}_v$. We subtract this divergence in the
$\overline{\rm MS}$ scheme and introduce the anomalous dimension 
$\gamma_v = \frac{{\rm d} \ln \tilde{Z}_v}{{\rm d} \ln \mu}$
which is given by
\begin{eqnarray}
  \gamma_v &=& - \frac{\alpha\alpha_s}{\pi^2 s_W^2}C_F
  \frac{\pi^2}{4} \frac{m_t^4}{M_W^2M_h^2} 
  \,.
\end{eqnarray}

In the regions $m_t\ll M_H$ and $m_t\approx M_H$ three, respectively,
six expansion terms have been evaluated. Our results read
\begin{eqnarray}
  c_{v,0}^{H, \rm mix} &=& \frac{m_t^2}{M_W^2} \Bigg [
  \left(
  \frac{\pi^2}{8}\ln{\frac{m_t^2}{\mu^2}}
  -\frac{29}{216}-\frac{277\pi^2}{2304}-\frac{\pi^2\ln{2}}{8}
  -\frac{21\zeta_3}{16}
  +\frac{139}{216}\ln\yh^2\right.  
  \nonumber  \\ &&\left. 
  -\frac{103}{288}\ln^2\yh^2
  \right)\yh^2 + \left(
  \frac{583}{576}-\frac{875\pi^2}{6912}+\frac{151}{192}\ln\yh^2
  -\frac{17}{16}\ln^2\yh^2\right)\yh^4 
  \nonumber \\ && 
  +\left(
  \frac{1533691}{432000}-\frac{27103\pi^2}{138240}-\frac{66647}{43200}\ln\yh^2
  -\frac{2251}{720}\ln^2\yh^2\right)\yh^6 + \ldots \Bigg ] 
  \,,\nonumber\\
  c_{v,1a}^{H, \rm mix} &=&  \frac{m_t^2}{M_W^2} \Bigg [
   \frac{\pi^2}{8} \frac{1}{1-\yhonea}
  \ln{\frac{m_t^2}{\mu^2}} 
  -5.760-5.533\yhonea
  \nonumber \\  && \mbox{}
  -5.704\yhonea^2-5.888\yhonea^3-6.053\yhonea^4 -
  6.200\yhonea^5 + \ldots \Bigg ] 
  \,,\nonumber\\
  c_{v,1b}^{H, \rm mix} &=&  \frac{m_t^2}{M_W^2} \Bigg [
  \frac{\pi^2}{8} \left ( 1-\yhoneb \right
  ) \ln{\frac{m_t^2}{\mu^2}} 
  -5.760 +5.533 \yhoneb   -0.171 \yhoneb^2
  \nonumber \\ && \mbox{}
  +0.0124\yhoneb^3+0.0304\yhoneb^4+0.0296 \yhoneb^5 +\ldots \Bigg ] 
  \,,
  \label{eq::cv2exp}
\end{eqnarray}
where $m_t$ is the on-shell mass and $\zeta_3\approx1.20206$ 
is Riemann's zeta-function.
The $\ln(m_t^2/\mu^2)$ term reminds on the divergence which has been
subtracted minimally.
Since in the limit $m_t\ll M_H$ some coefficients of the
$\epsilon$-expansion of the master integrals could only be computed
numerically the results for the finite parts of $c_{v,1a}^{H, \rm
  mix}$ and $c_{v,1b}^{H, \rm mix}$ are presented in numerical form.
We also want to stress again that $c_{v,1a}^{H, \rm mix}$ and
$c_{v,1b}^{H, \rm mix}$ contain the same information. 
However, expressed in terms of $\yhoneb$ the convergence
properties are much better.

\begin{figure}[t]
  \begin{center}
    \begin{tabular}{c}
      \epsfig{figure=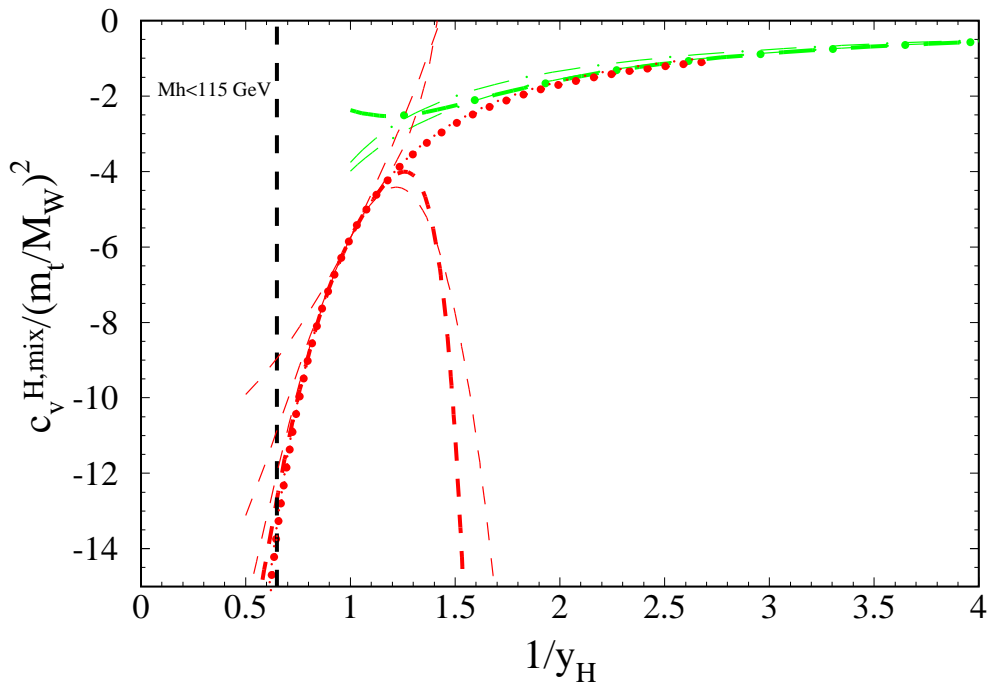,width=35em}
    \end{tabular}
    \parbox{14.cm}{
      \caption[]{\label{fig::cv2l}\sloppy
        $c_{v}^{\rm mix}$ as a function of $1/\yh$.
        The approximations are shown as 
        dotted, dashed and dash-dotted lines where for the expansion
        around $\yh\approx 1$ two different parameterizations have been
        chosen. Lower-order terms are plotted using thin lines.
	For the renormalization scale $\mu=m_t$ has been adopted.
        }}
  \end{center}
\end{figure}

In Fig.~\ref{fig::cv2l} the results are shown where the same notation as
for the one-loop result has been adopted and $\mu=m_t$ has been
chosen. Next to the highest expansion
terms we also include the lower-order terms as thin lines which
nicely demonstrates the convergence properties in the individual regions.
Comparing the two parameterizations of the expansion around $M_t=M_h$
one observes that the one in terms of $\yhoneb$ demonstrates a much
better convergence behaviour: the thin dots in Fig.~\ref{fig::cv2l}
are practically indistinguishable from the fat ones. This can also be
seen in Eq.~(\ref{eq::cv2exp}) where the coefficients become quite
small starting from the third term which is not the case for
$c_{v,1a}^{H, \rm mix}$.
Thus, the combination of $c_{v,1b}^{H, \rm mix}$ and $c_{v,0}^{H, \rm mix}$
provide a very good approximation to $c_v^H$
for Higgs boson masses above $M_h=115$~GeV.

\section{\label{sec::appl}Phenomenological application}

Let us in a first step discuss the effect on the top quark mass. The
connection between the position of the peak in the threshold cross
section, $E_{\rm res}$, and the top quark mass is given by
\begin{eqnarray}
  E_{\rm res} &=& 2m_t + E_1^{\rm p.t.} + \delta^{\Gamma_t} E_{\rm res}
  \,,
  \label{eq::Eres}
\end{eqnarray}
where $E_1^{\rm p.t.}$ is the perturbative part of the ground state energy
and $\delta^{\Gamma_t} E_{\rm res}=100\pm10$~MeV~\cite{Penin:2002zv} 
takes into account the effect of
the finite width, the higher order resonances and the continuum. 
Non-perturbative
effects are negligible for the top quark system.
$E_1^{\rm p.t.}$ up to third order in QCD has been computed in
Ref.~\cite{Penin:2002zv}. The contribution from the Higgs boson is
given in Eq.~(\ref{eq::E_n^H}) where $\alpha_s$ has to be evaluated at
the typical soft scale given by $\mu_s=C_F\alpha_s(\mu_s) m_t\approx 30$~GeV.

According to Eq.~(\ref{eq::Eres}) a shift in
$E_1^{\rm p.t.}$ can directly be translated into a shift in the top
quark mass, $\delta m_t$.
Choosing for the input values $m_t=175$~GeV, $M_W=80.41$~GeV, $s_W^2=0.23$,
$\alpha=1/127$~\cite{Jegerlehner:2001ca}, $\alpha_s(M_Z)=0.118$,
$M_H=120$~GeV and $\mu_s=30$~GeV we observe for 
$\delta m_t \approx - E_1^H/2 \approx 26$~MeV.
This reduces to $9(1)$~MeV for $M_H=200(500)$~GeV.
If $M_H=120$~GeV is adopted and $\mu_s=15(60)$~GeV is chosen 
one obtain $\delta m_t \approx 38(18)$~MeV.
Thus, for light Higgs masses
relatively large effects are observed whereas for larger $M_H$ the 
numerical effect on $m_t$ is small.
Note that our findings are in agreement with the more qualitative
analysis of Ref.~\cite{Strassler:1990nw}.

As a second application it is interesting to consider the effect of
the new corrections to the normalized cross section
$R=\sigma(e^+e^-\to t\bar t)/\sigma(e^+e^-\to\mu^+\mu^-)$ at the resonance
energy which is dominated by the contribution from the would-be toponium
ground-state. It is of the form
\begin{eqnarray}
  R_1&=&R_1^{\rm LO} c_v^2(m_t) \frac{|\psi_1(0)|^2}{|\psi_1^C(0)|^2}
  +\ldots
  \label{eq::R}
\end{eqnarray}
with
$R_1^{\rm LO}=6\pi N_cQ_t^2|\psi_1^C(0)|^2/\left(m_t^2\Gamma_t\right)$.
The ellipses in Eq.~(\ref{eq::R}) denote contributions from higher
order operators.
Note that the divergences in $c_v$ and $\psi_1(0)$ cancel in the
combination of Eq.~(\ref{eq::R}).
The most advanced evaluation of $R_1$ is provided in Ref.~\cite{Penin:2005eu}
where all logarithmically enhanced third-order corrections 
and the ones proportional to $\beta_0^3$ are included. Let us for
completeness repeat the final numerical result which is given by
\begin{eqnarray}
  R_1 &\approx&
  R_1^{\rm LO}(1 -0.244_{\rm NLO}
  +0.449_{\rm NNLO}-0.277_{\rm N^3LO^\prime}
  + \delta_H^{(1)} + \delta_H^{(2)} \ldots)
  \,,
  \label{eq::R1}
\end{eqnarray}
where $\mu_s=30$~GeV has been chosen.
The prime reminds that the N$^3$LO corrections are not complete
and $\delta_H^{(1)}$ and $\delta_H^{(2)}$ parameterize the one- and two-loop
corrections due to the Higgs boson considered in the present paper. 
Using $M_H=120/200/500$~GeV and
$\mu_s=30$~GeV we obtain $\delta_H^{(1)}=0.067/0.034/0.009$ and
$\delta_H^{(2)}=0.036/0.011/0.0002$. 
Thus, moderate effects are observed for Higgs boson masses below 
about 200~GeV. However, in contrast to the pure QCD effects of
Eq.~(\ref{eq::R1}) the convergence seems to be much better as 
$\delta_H^{(2)}$ it is substantially smaller than $\delta_H^{(1)}$.

In principle it is possible to apply the formulae of this Section also
to the bottom system. However, due to the suppression factor
$m_b^2/M_H^2$ the numerical effect is very small for Higgs boson
masses above 100~GeV.


\section{\label{sec::concl}Conclusions}

In this paper we take an important step towards the evaluation of the mixed
QCD/electroweak corrections to the threshold production of top quark
pairs, namely, the complete Higgs boson mass dependence is computed.
In particular, the two-loop corrections to the matching coefficients
are evaluated and the corresponding operator is introduced in the
effective theory.
Moderate numerical effects on the top quark mass and the peak cross
section are observed for Higgs boson masses below 200~GeV.
Considering the anticipated precision of an ILC it is certainly
necessary to take into account these corrections.


\vspace*{1em}

\noindent
{\large\bf Acknowledgements}\\
We would like to thanks J.H. K\"uhn, 
A.A. Penin, V.A. Smirnov and O.V. Tarasov
for useful discussions and A.A. Penin for
carefully reading the manuscript.
This work was supported by the ``Impuls- und Vernetzungsfonds'' of the
Helmholtz Association, contract number VH-NG-008 and the SFB/TR~9.


\begin{appendix}


\section{\label{app::Z-boson}Matching coefficients for the $Z$ boson
  exchange diagrams}

\begin{figure}[ht]
  \begin{center}
    \begin{tabular}{cc}
      \epsfig{figure=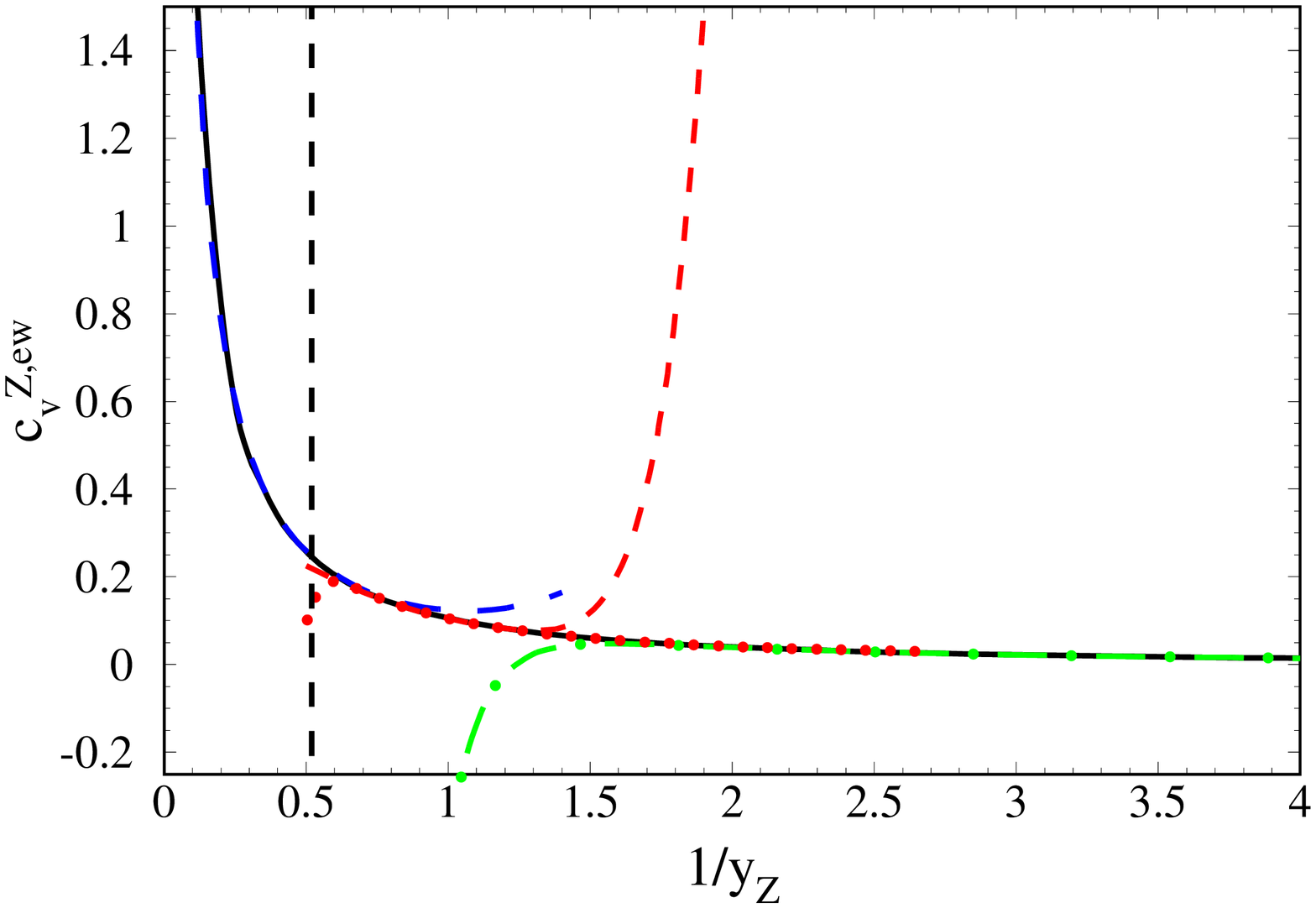,width=18em}
      &
      \epsfig{figure=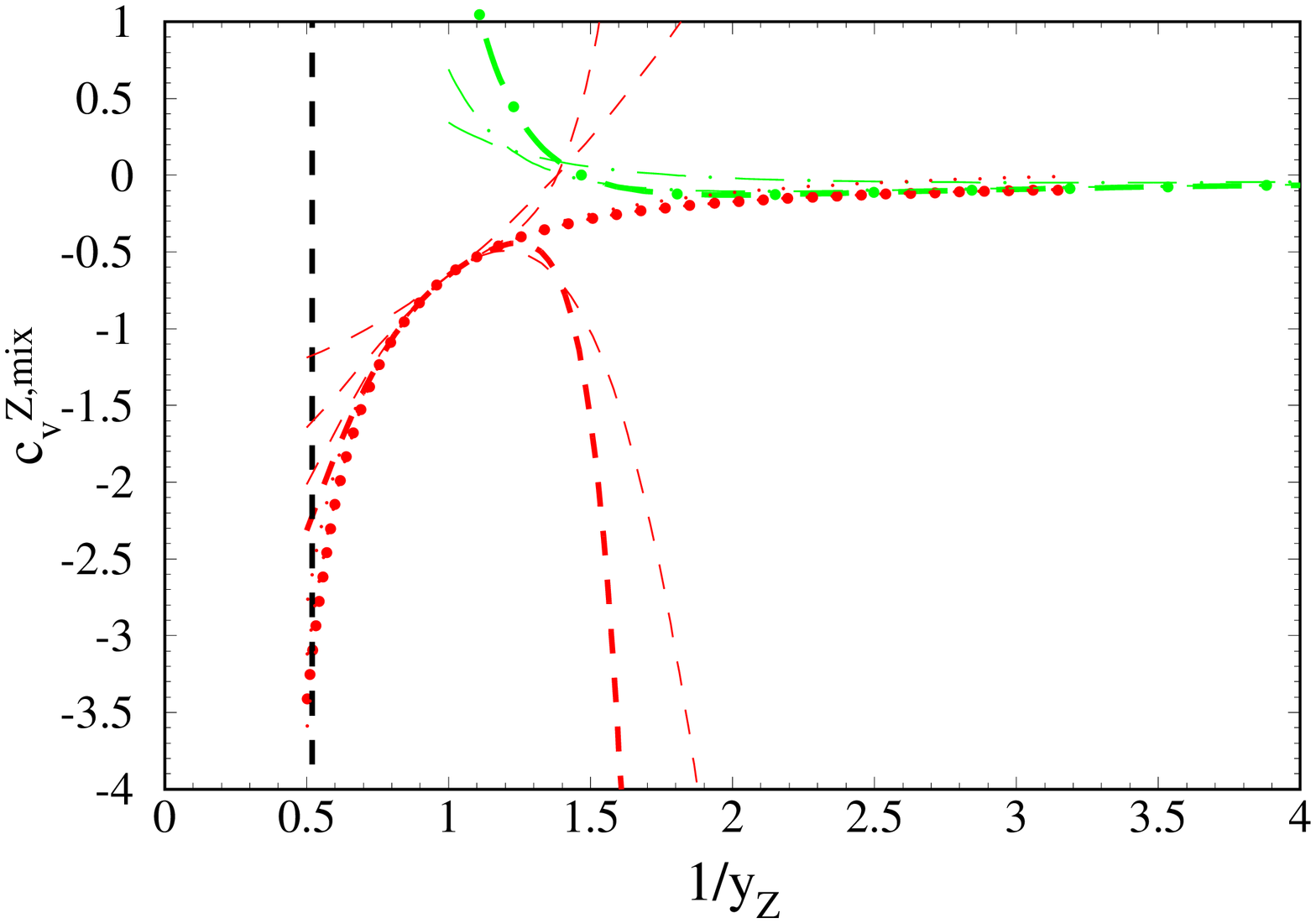,width=18em}
      \\
      (a) & (b)
    \end{tabular}
    \parbox{14.cm}{
      \caption[]{\label{fig::Z}\sloppy
        (a) $c_{v}^{Z, \rm ew}$ and (b) $c_{v}^{Z, \rm mix}$
        as a function of $1/\yz$.
        The same notation as in Figs.~\ref{fig::cv1l}
        and~\ref{fig::cv2l} is adopted.
        The vertical line indicates the physical value of $1/\yz=M_Z/m_t$.
        }}
  \end{center}
\end{figure}

In the following we present the results for the matching coefficients
induced by vertex corrections involving the $Z$ boson and the
corresponding Goldstone boson, respectively.
The calculation is similar to the diagrams involving a Higgs boson
which is presented in the main part of the paper. Note that there are
also box diagrams involving the $Z$ boson which are not considered here.
The one-loop result is given by~\cite{Guth:1991ab}
\begin{eqnarray}
   c_v^{Z, \rm ew} 
  &=&  a_t^2 s_W^2 \Bigg[ \frac{5\yz^2-2}{12\yz^2} 
    - \frac{3\yz^4-4\yz^2+1}{12\yz^4} \ln\yz^2 +
   \frac{(\yz^2-1)^2}{6\yz^2}  \Psi(\yz) \Bigg] 
  \nonumber \\ && 
  \!\!\!
  + v_t^2 s_W^2 \Bigg[
    -\frac{3\yz^2+2}{12\yz^2} - \frac{1}{12\yz^4} \ln\yz^2 +
    \left(\frac{7}{48}+\frac{1}{6\yz^2}+\frac{\yz^2}{4}-\frac{3}{16(4\yz^2-1)}
    \right) \Psi(\yz) 
    \Bigg]\,,
  \nonumber\\
  \label{eq::cv1zexact}
\end{eqnarray}
with 
$\yz=m_t/M_z$, $a_t=1/(2s_Wc_W)$, $v_t=(1/2- 4 s_W^2/3)/(s_Wc_W)$ and
$c_W = \sqrt{1-s_W^2} = M_W/M_Z$. The function $\Psi(x)$ can be found in
Eq.~(\ref{eq::psi}). It is straightforward to obtain the expansion in
the various kinematical regions. Thus we refrain from listing them
explicitly. 
In Fig.~\ref{fig::Z}(a) the exact result is plotted together with the
expansions where in each region the same depth is chosen as in 
Eq.~(\ref{eq::cv1exp}).
The vertical dashed line marks the phenomenological result for $\yz$
which is nicely approximated both from the $m_t\gg M_Z$ and 
the $m_t\approx M_Z$ region. 
However, it should be noted that both parameterizations for 
the $m_t=M_h$ expansion become instable below $1/y_Z\approx 0.6$.

At two-loop order we consider the diagrams of
Fig.~\ref{fig::diags} where the Higgs boson is replaced by the $Z$
boson. The reduction of the occuring integrals is in close analogy and
in fact the same set of master integrals is necessary.
Although the region $m_t\ll M_Z$ is phenomenologically not relevant we
nevertheless present the results since it constitutes an important
cross checks for the other kinematical limits. Furthermore, it is
possible to obtain from this limit the result for the case of the
bottom quark.
We adopt the notation from Eq.~(\ref{eq::cvdef}) and denote the
contribution from the $Z$ boson to $c_v^{\rm mix}$ by 
$c_{v}^{Z, \rm mix}$. The additional subscript ``0'', ``1a'' or ``1b''
reminds on the kinematical region considered.
Our results read
\begin{eqnarray}
  c_{v,0}^{Z, \rm mix} &=& 
  \left( v_t^2+a_t^2 \right) s_W^2 \frac{\pi^2}{8}\ln \frac{m_t^2}{\mu^2} 
    \nonumber \\ && 
   + v_t^2 s_W^2 \Bigg [ \left(
    - \frac{587}{432} +\frac{131\pi^2}{576}
    -\frac{\pi^2\ln{2}}{8}  
    -\frac{21}{16}\zeta_3
    +\frac{28}{27} \ln\yz^2
    -\frac{11}{72}\ln^2\yz^2
    \right ) \yz^2 
    \nonumber \\ && 
    +\left(
    -\frac{403}{64}+\frac{205\pi^2}{576} 
    + \frac{2299}{864}\ln\yz^2-\frac{5}{18}\ln^2\yz^2  
    \right )\yz^4  
    \nonumber \\ && 
    +\left(
    - \frac{21091951}{864000} 
    + \frac{3951\pi^2}{2560} 
    + \frac{28301}{7200}\ln\yz^2
    + \frac{469}{160}\ln^2\yz^2
    \right ) \yz^6 
    + \ldots 
    \Bigg]
  \nonumber \\ && 
  + a_t^2 s_W^2\Bigg[ \left(
    -\frac{281}{432} +\frac{85\pi^2}{192}
    -\frac{\pi^2\ln{2}}{8}  
    -\frac{21}{16}\zeta_3
    +\frac{29}{54} \ln\yz^2
    -\frac{11}{72}\ln^2\yz^2
    \right ) \yz^2 
    \nonumber \\ && 
    +\left(
    \frac{709}{576}+\frac{11\pi^2}{432} 
    + \frac{521}{288}\ln\yz^2-\frac{79}{72}\ln^2\yz^2  
    \right )\yz^4 
    \nonumber \\ && 
    +\left(
      \frac{260687}{32000} 
    - \frac{16013\pi^2}{69120}
    + \frac{32453}{21600}\ln\yz^2
    - \frac{5479}{1440} \ln^2\yz^2
    \right ) \yz^6 
    + \ldots\Bigg]
  \,,\nonumber
\end{eqnarray}
\begin{eqnarray}
  c_{v,1a}^{Z, \rm mix} &=&  
  \left( v_t^2+a_t^2 \right) s_W^2 \frac{\pi^2}{8}\ln \frac{m_t^2}{\mu^2} 
  \frac{1}{1-\yzonea} 
  \nonumber\\&&\mbox{}
  + v_t^2 s_W^2\Bigg [ 
    -4.564 - 4.699 \yzonea - 5.009 \yzonea^2 
    - 5.277 \yzonea^3 
    - 5.502 \yzonea^4 
    \nonumber \\ &&
    - 5.693 \yzonea^5 + \ldots \Bigg ] 
  + a_t^2 s_W^2\Bigg [
    - 1.324 
    - 1.504 \yzonea 
    - 1.740 \yzonea^2 
    \nonumber \\ &&
    - 1.930 \yzonea^3 
    - 2.083 \yzonea^4 
    - 2.212 \yzonea^5 + \ldots  \Bigg ] 
  \,,\nonumber
\end{eqnarray}
\begin{eqnarray}
  c_{v,1b}^{Z, \rm mix} &=&  
  \left( v_t^2+a_t^2 \right) s_W^2 \frac{\pi^2}{8}\ln\frac{m_t^2}{\mu^2} 
  \left(1-\yzoneb\right)
  + v_t^2 s_W^2\Bigg [ 
    - 4.564 
    + 4.699 \yzoneb 
    \nonumber \\ &&
    - 0.311 \yzoneb^2 
    - 0.043 \yzoneb^3 
    + 0.001 \yzoneb^4 
    + 0.011 \yzoneb^5 + \ldots \Bigg ] 
  \nonumber \\ &&
  + a_t^2 s_W^2\Bigg [ 
    - 1.324 
    + 1.504 \yzoneb 
    - 0.236 \yzoneb^2 
    - 0.047 \yzoneb^3 
    - 0.011 \yzoneb^4 
  \nonumber \\ &&
  - 0.001 \yzoneb^5 + \ldots  \Bigg ] 
  \,,
  \label{eq::cv2zexp}
\end{eqnarray}
where the divergence has been subtracted in a minimal way and
$\yz=m_t/M_Z$, $\yzonea = (1-1/\yz^2)$ and $\yzoneb = (1-\yz^2)$.
In Fig.~\ref{fig::Z}(b) the results of Eq.~(\ref{eq::cv2zexp}) are shown
as a function in $1/\yz$. Similarly to the two-loop case one observes
a rapid convergence for $c_{v,1b}^{Z, \rm mix}$ whereas the magnitude
of the coefficients of the higher order terms in
$c_{v,1a}^{Z, \rm mix}$ even increase.
This gives us quite some confidence that $c_{v,1b}^{Z, \rm mix}$
evaluated at the physical scale provides a very good approximation to
the unknown exact result. Using $1/\yz=91.19/175\approx0.52$ we get
\begin{eqnarray}
  c_v^{Z,\rm mix}\Bigg|_{\mu^2=m_t^2} &\approx& -3.1 \pm 0.3
  \,,
  \label{eq::cvZnum}
\end{eqnarray}
where we assign a generous uncertainty of 10\% which results from the
comparison of lower-order approximations.
The accuracy of the 
result for $c_v^{Z,\rm mix}$ as given in Eq.~(\ref{eq::cvZnum})
is more than sufficient as far as the forseen precision of the
measurement of the threshold top quark production is concerned.


\section{\label{app::z2}Addendum to Ref.~\cite{Eiras:2005yt}}

\begin{figure}[t]
  \begin{center}
    \begin{tabular}{cc}
      \epsfig{figure=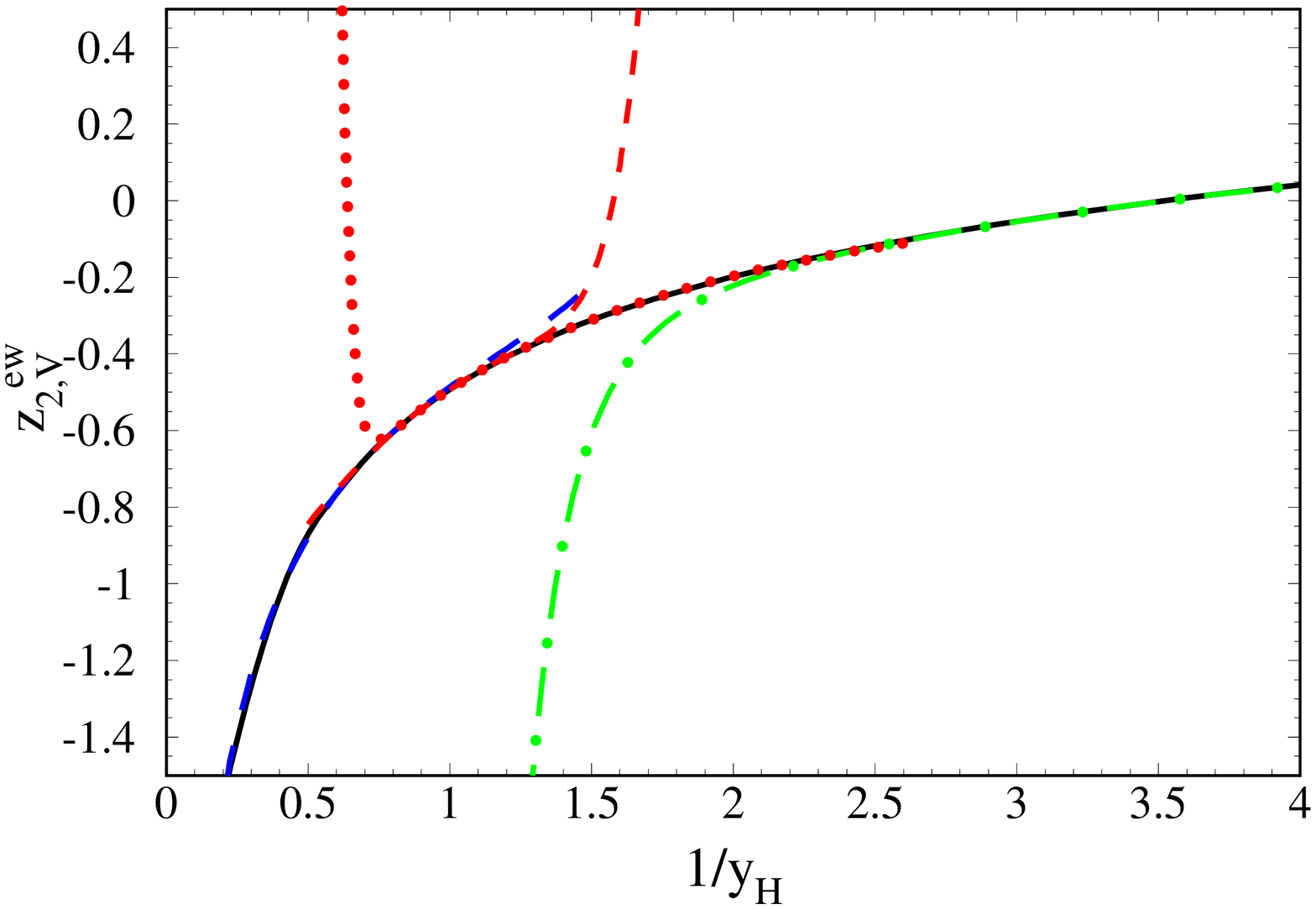,width=18em}
      &
      \epsfig{figure=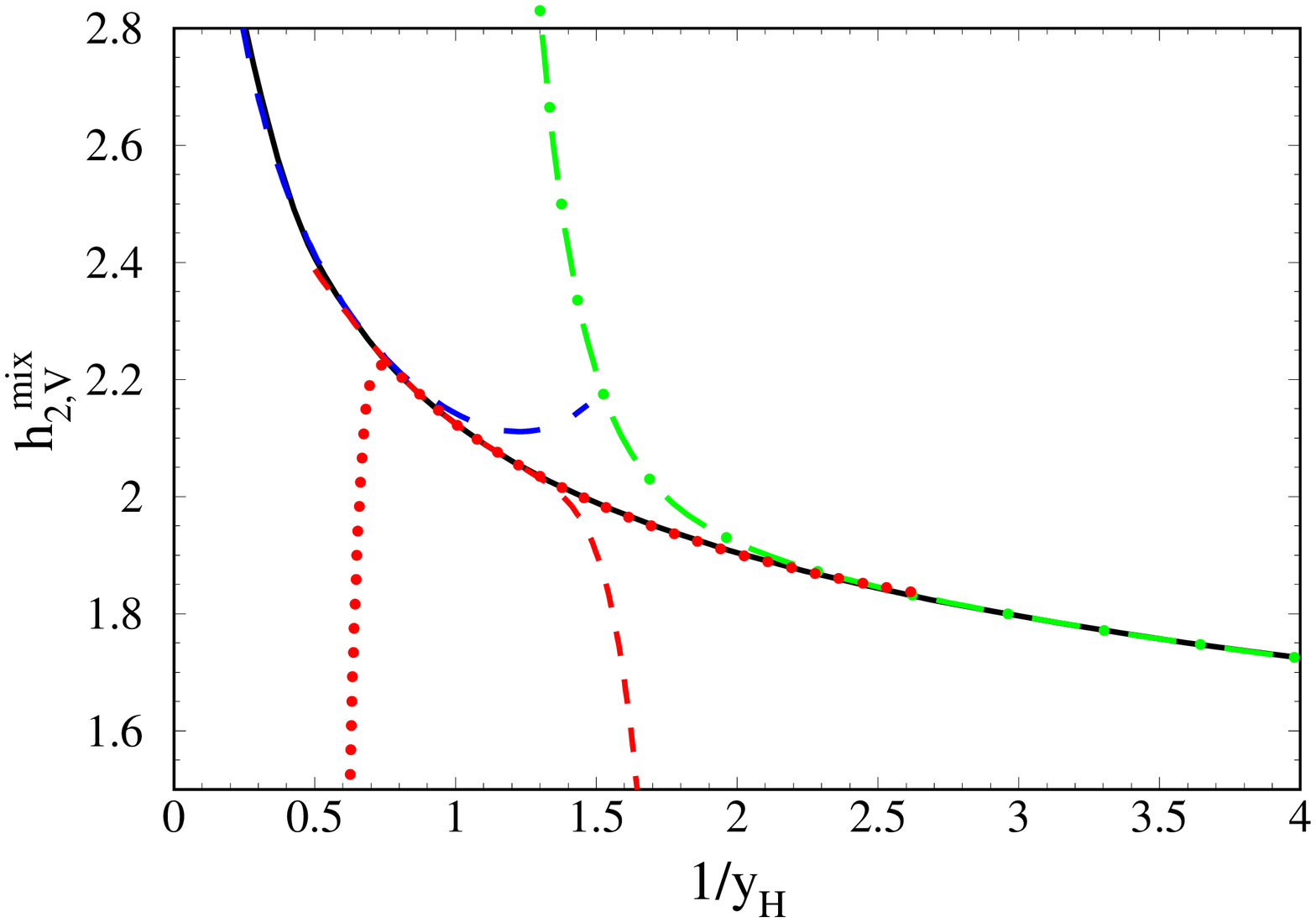,width=18em}
      \\(a) & (b)\\[-1.5em]
      \multicolumn{2}{c}{
        \epsfig{figure=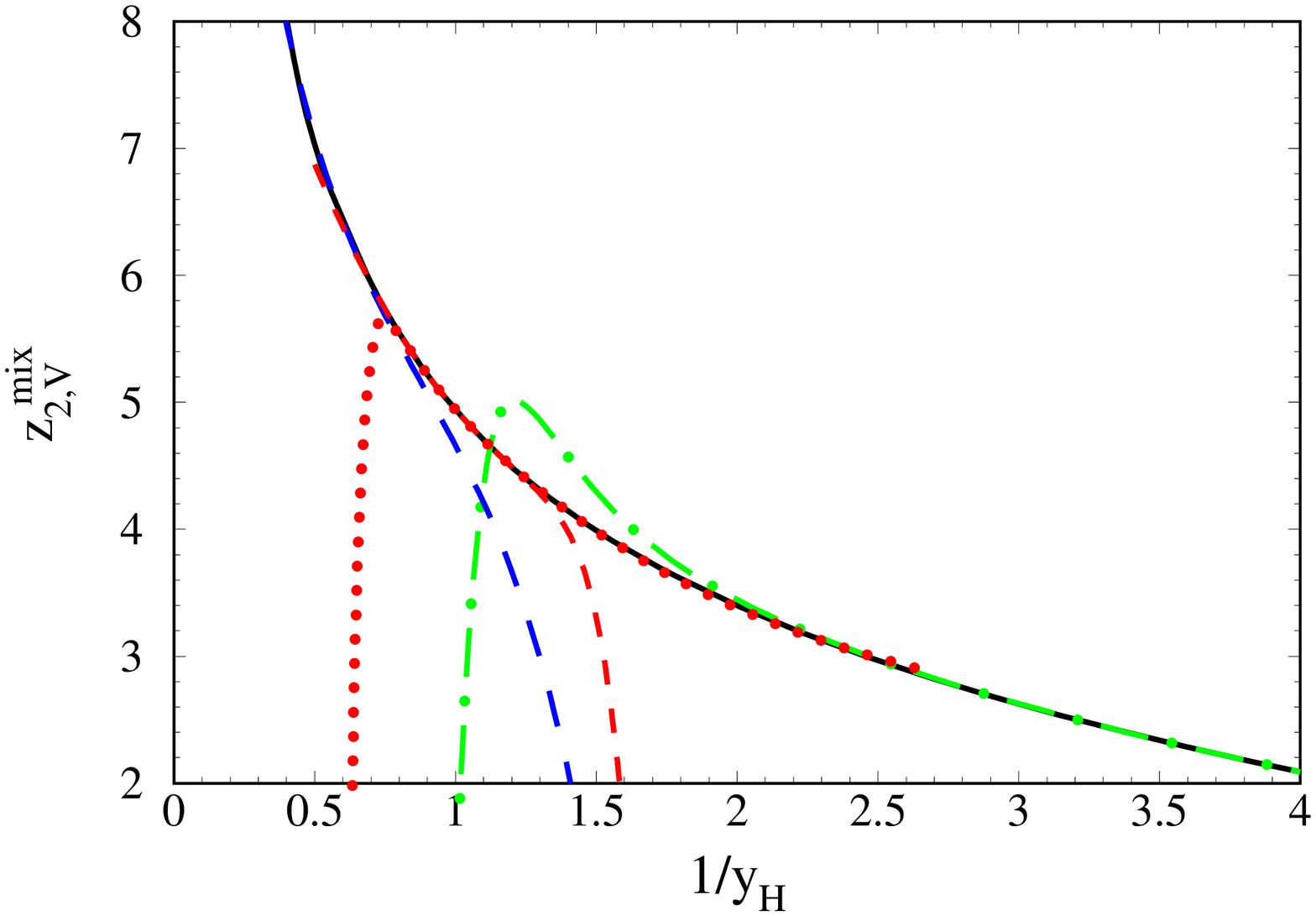,width=18em}
      }
      \\ \multicolumn{2}{c}{(c)}
    \end{tabular}
    \mbox{}
    \parbox{14.cm}{
      \caption[]{\label{fig::z2v_2}\sloppy
        (a) One-loop, (b) $1/\epsilon$ pole and (c) constant part of 
        the two-loop corrections to $Z_2^{V, \rm OS}$ as a
        function of $1/y_H=M_H/m_t$.
        The solid (black) line represents the exact result.
        The approximations in the three regions are shown as 
        dotted, dashed and dash-dotted lines where for the expansion
        around $\yh\approx 1$ two different parameterizations (short
        dashes and dots) have been chosen.}}
  \end{center}
\end{figure}

In this Appendix we want to provide an update of the results
of Ref.~\cite{Eiras:2005yt}. In Fig.~4 of that reference
the Higgs boson mass dependence of the on-shell wave function
renormalization constant to one- and two-loop order has been
plotted as a function of $1/\yh = M_H/m_t$. Next to the exact 
results also the ones for 
$m_t\ll M_H$, $m_t\approx M_H$ and $m_t\gg M_H$
have been shown and good agreement over almost the whole range in
$\yh$ has been found. However, there was a small gap around 
$1/\yh\approx 2$ which was not covered very well. This deficit is 
removed in Fig.~\ref{fig::z2v_2} where the result obtained in the limit
$m_t\approx M_H$ is plotted both in terms of $\yhonea = (1-1/\yh^2)$
(dashed) and $\yhoneb = (1-\yh^2)$ (dotted). 
One can see that the former is valid down
to quite small values of $1/\yh$ whereas the validity of the latter 
expansion reaches up to larger values of $1/\yh$ leading to
a significant overlap with the expansions for large and small 
values of $\yh$.
Thus, in the whole $\yh$ range the simple expansions approximate the
exact result to a very high precision.

Let us also mention that there is a misprint in the definition of $v_t$ in 
Ref.~\cite{Eiras:2005yt} after Eq.~(17): it is too small by a factor two.
Furthermore, the analytic expression for $Y^\epsilon$ in Eq.~(22) has
to be multiplied by ($-1$) and in Eq.~(27) a factor $1/s_W^2$ is missing
in the terms containing $\alpha_s/\pi$.


\end{appendix}



\end{document}